\documentclass{eas}
\usepackage{graphicx}
\begin{document}

\title{Hunting for the signatures of molecular cloud formation} 
\author{Simon C. O. Glover}\address{Universit\"{a}t Heidelberg, Zentrum f\"{u}r Astronomie, Institut f\"{u}r Theoretische Astrophysik, Albert-Ueberle-Stra{\ss}e 2, 69120 Heidelberg, Germany; \email{glover@uni-heidelberg.de}}
\author{Paul C. Clark}\address{School of Physics and Astronomy, Cardiff University, 5 The Parade, Cardiff CF24 3AA, UK; \email{paul.clark@astro.cf.ac.uk}}
\begin{abstract}
In order to understand how molecular clouds form in the Galactic interstellar medium, we would like to be able to map the structure and kinematics of the gas flows responsible for forming them. However, doing so is observationally challenging. CO, the workhorse molecule for studies of molecular clouds, traces only relatively dense gas and hence only allows us to study those portions of the clouds that have already assembled. Numerical simulations suggest that the inflowing gas that forms these clouds is largely composed of CO-dark H$_{2}$. These same simulations allow us to explore the usefulness of different tracers of this CO-dark molecular material, and we use them here to show that the [C$\,${\sc ii}] fine structure line is potentially a very powerful tracer of this gas and should be readily detectable using modern instrumentation.
\end{abstract}
\maketitle
\section{Introduction}
Molecular clouds are a key component of the interstellar medium (ISM), as they represent the locations where all star formation takes place. For this reason it is very important to understand how molecular clouds form. Presently, the leading theory is that molecular clouds form from converging flows of lower-density gas, driven either by stellar feedback or by large-scale gravitational instability (see Dobbs \etal, \cite{dobbs14} or Klessen \& Glover \cite{kg14} for some recent reviews of the different possible formation mechanisms). In this contribution, we briefly discuss one of the main difficulties involved in observationally distinguishing between these different models, and how we might hope to overcome it.

\section{When do molecular clouds become molecular?}
From a chemical point of view, the main obstacle standing in the way of a given patch of cold atomic gas becoming molecular is the influence of the interstellar radiation field (ISRF). The H$_{2}$ formation rate per unit volume in the cold neutral medium (CNM) can be written approximately as (Hollenbach \& McKee \cite{hm79})
\begin{equation}
R_{\rm form} \simeq 1.5 \times 10^{-17} n n_{\rm H} \: {\rm cm^{-3} \, s^{-1}},
\end{equation}
where $n_{\rm H}$ is the number density of H atoms and $n$ is the number density of H nuclei. In the absence of shielding, the H$_{2}$ photodissociation rate per unit volume is given by (Draine \& Bertoldi \cite{db96})
\begin{equation}
R_{\rm pd} = 5.6 \times 10^{-11} n_{\rm H_{2}} \: {\rm cm^{-3} \, s^{-1}},
\end{equation}
where $n_{\rm H_{2}}$ is the number density of H$_{2}$ molecules. For densities typical of those found in the CNM, $n \sim 10$--$100 \, {\rm cm^{-3}}$, the resulting equilibrium ratio of H$_{2}$ to H is very small, $n_{\rm H_{2}}/n_{\rm H} \sim 10^{-6}$. As H$_{2}$ is a necessary pre-requisite for CO formation, and CO photodissociation by the ISRF is also highly efficient, the fractional abundance of CO in these conditions is also very small. 

Consequently, the gas can only become molecular once it is capable of shielding itself from the ISRF. In the case of H$_{2}$, self-shielding dominates when the radiation field is weak, becoming effective once the hydrogen column density exceeds $N_{\rm H} \sim 2 \times 10^{20} \: {\rm cm^{-2}}$. In the case of CO, however, shielding by dust dominates, becoming effective only at $N_{\rm H} > 4 \times 10^{21} \: {\rm cm^{-2}}$. There is thus a range of column densities for which the gas will be H$_{2}$-rich but CO-poor. This CO-dark molecular gas has attracted increasing interest in recent years (see e.g.\ Wolfire \etal\ 2010, Langer \etal\ 2014), particularly since simulations of converging flows have demonstrated that much of the inflowing gas will be CO-dark H$_{2}$ (Heitsch \& Hartmann \cite{hh08}, Clark \etal\ \cite{clark12}). Because of its chemical composition, this gas cannot properly be traced using either H$\,${\sc i} or CO observations. However, it may be possible to observe this gas using the  [C$\,${\sc ii}] $158 \: \mu$m fine structure line. We explore the feasibility of this in the next section.

\section{Using [CII] emission to probe CO-dark molecular gas}
\label{CII}
\begin{figure}
\begin{center}
\includegraphics[width=0.92\textwidth]{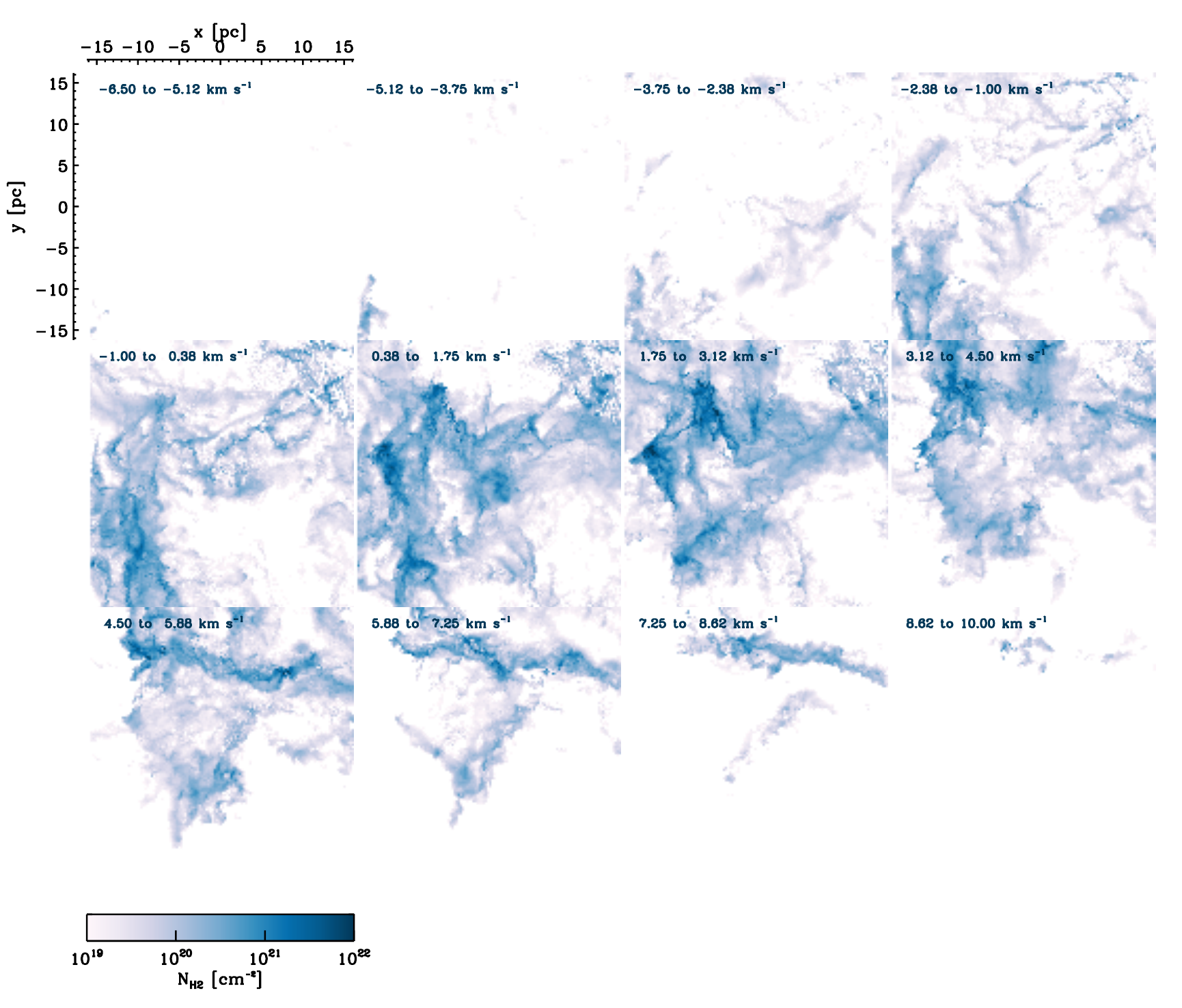}
\caption{H$_{2}$ column density in each of the indicated velocity channels for one of the
molecular clouds formed in the simulation described in Section~\ref{CII}. Each velocity
channel has a width of $1.375 \, {\rm km \, s^{-1}}$. \label{H2chan}}
\end{center}
\end{figure}

\begin{figure}
\begin{center}
\includegraphics[width=0.92\textwidth]{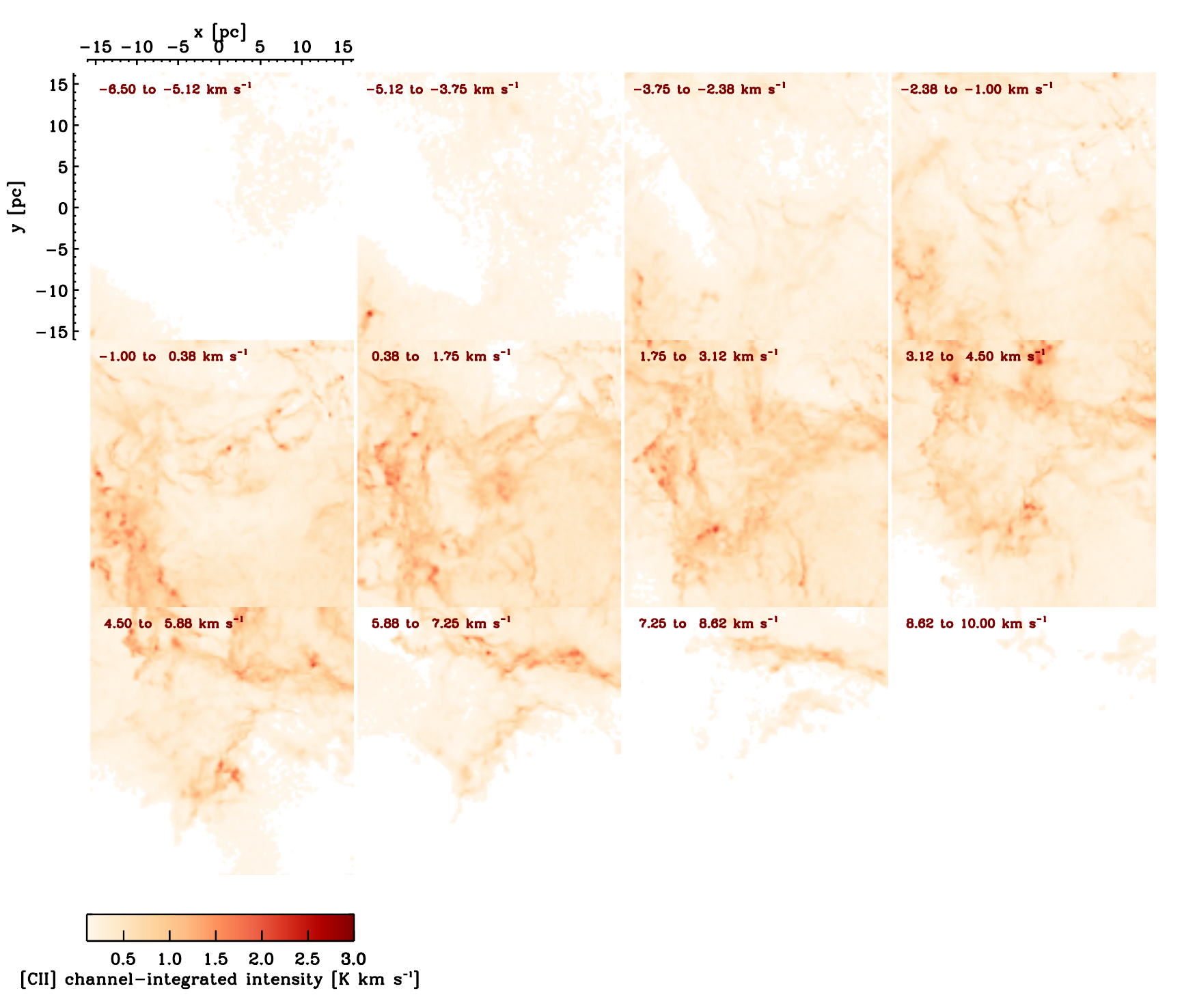}
\caption{As Figure~\ref{H2chan}, but for the  [C$\,${\sc ii}]  channel-integrated intensities. \label{CIIchan}}
\end{center}
\end{figure}

To assess how easily we can trace molecular cloud assembly using [C$\,${\sc ii}] emission, we have carried out a simulation of a representative (200~pc)$^{3}$ volume of the ISM using the {\sc arepo} moving mesh code (Springel \cite{springel10}). We inject turbulence on large scales and follow the resulting chemical, thermal and dynamical evolution of the gas using the chemistry and cooling module described in Smith \etal\ (\cite{smith14}). We assume that the gas is illuminated by a uniform ISRF, and have carried out simulations with  ISRF strengths of $G_{0} = 1$ and 10 in Habing (\cite{habing68}) units. The attenuation of the ISRF within the dense clouds that form within the simulations is modelled using the {\sc treecol} algorithm (Clark \etal\ \cite{clark12b}). 

A number of molecular clouds form in each simulation. We have selected a subset of these and post-processed them using the {\sc radmc-3d} radiative transfer code to produce synthetic emission maps of the [C$\,${\sc ii}] $158 \: \mu$m line. We show some example results in Figures~\ref{H2chan} and \ref{CIIchan}. Figure~\ref{H2chan} shows the H$_{2}$ column density through one of the clouds from our $G_{0} = 10$ simulation, broken up into a series of velocity channels of width $1.375 \: {\rm km \, s^{-1}}$. Figure~\ref{CIIchan} shows the [C$\,${\sc ii}] integrated intensity in the same set of velocity channels. Comparison of the two figures shows clearly that the [C$\,${\sc ii}] line emission traces the full velocity width of the molecular gas, with little contamination from the surrounding atomic gas. In addition, it is clear that the peak [C$\,${\sc ii}] brightness temperatures are 
typically $\sim$1--2~K over a large area. This is important, as this is bright enough to be mapped with high velocity resolution over angular areas of a few square arc-minutes using the {\it upGREAT} instrument on board the SOFIA observatory in a matter of hours (S.~Ragan, priv.\ comm.), making it plausible to contemplate making such maps of real clouds. Our $G_{0} = 1$ simulation yields similar results, but in this case the diffuse CO-dark H$_{2}$ is cooler and the resulting [C$\,${\sc ii}] emission is considerably weaker.


\end{document}